# Categorize radio interference using component and temporal analysis

Mao Yuan ,[1,2]★ Weiwei Zhu,[1]★ Haiyan Zhang,[1,3] Shijie Huang,[1] Mengyao Xue ,[1] Di Li ,[1]
Youling Yue,[1] Pei Wang,[1] Jiarui Niu,[1,2] Yuxuan Hu ,[4] Chunjiang Li,[4] Chenchen Miao ,[1,2]
Yu Wang ,[1,2] Lingqi Meng[1,2] and Bo Peng[1]★

[1]*National Astronomical Observatories of the Chinese Academy of Sciences (NAOC), Beijing 100101, China*
[2]*College of Astronomy and Space Sciences, University of Chinese Academy of Sciences, Beijing 100049, China*
[3]*Hebei Key Laboratory of Radio Astronomy Technology, Hebei 050081, China*
[4]*Department of Astronomy, School of Physics, Peking University, Beijing 100101, China*

## ABSTRACT

Radio frequency interference (RFI) is a significant challenge faced by today's radio astronomers. While most past efforts were devoted to cleaning the RFI from the data, we develop a novel method for categorizing and cataloguing RFI for forensic purpose. We present a classifier that categorizes RFI into different types based on features extracted using Principal Component Analysis (PCA) and Fourier analysis. The classifier can identify narrowband non-periodic RFI above $2\sigma$, narrowband periodic RFI above $3\sigma$, and wideband impulsive RFI above $5\sigma$ with F1 scores [defined as $F1 = (2 \cdot recall \times precision)/(recall + precision)$] between 0.87 and 0.91 in simulation. This classifier could be used to identify the sources of RFI as well as to clean RFI contamination (particularly in pulsar search). In the long-term analysis of the categorized RFI, we found a special type of drifting periodic RFI that is detrimental to pulsar search. We also found pieces of evidence of an increased rate of impulsive RFI when the telescope is pointing towards the cities. These results demonstrate this classifier's potential as a forensic tool for RFI environment monitoring of radio telescopes.

## 1 INTRODUCTION

In recent decades, rising human activities have substantially increased Radio Frequency Interference (RFI) in astronomical observations (Njoku et al. 2005). Most of the astronomical radio signals are fainter than RFI. Scientists take many measures to prevent RFI from entering the receiver including setting up a radio-quiet zone and shielding radio transmitters, etc. Nevertheless, much undesired interference still gets through. While many efforts have been invented to the cleaning of the RFI from radio signals in the literature, this paper describes methods for categorizing and studying the RFI themselves as well as masking contaminated data.

Many general methods were designed to detect and mask RFI in radio astronomy, including rejection of time-domain impulses and flagging of frequency-domain bad channels (Fridman & Baan 2001; Guner, Johnson & Niamsuwan 2007; Offringa et al. 2010; Baan 2011; Tarongi & Camps 2011; Ford & Buch 2014; An et al. 2017). In specific fields, like pulsar searches, radio interference can be particularly detrimental (Eatough, Keane & Lyne 2009). First, RFI may mask and reduce the detected signal-to-noise ratio (SNR) of a potential new pulsar. Secondly, recurrent RFI, especially periodic ones like alternating current in power line interference, could generate many fake pulsar candidates that outnumber and mask real signals. Many techniques were invented to mitigate the impact of RFI in pulsar search. For example, PRESTO[1] is a widely used pulsar search software suite. It provides an executable program to detect and mask RFI called *rfifind* (Ransom, Eikenberry & Middleditch 2002). *rfifind* utilizes the maximum power in the Fourier domain, mean value, and variance of small blocks of data to flag and mask RFI based on preset thresholds. *rfifind* is generally very useful for masking impulsive and periodic RFI. Spatial filtering (Leshem, van der Veen & Boonstra 2000; Raza, Boonstra & Van der Veen 2002; Kocz, Briggs & Reynolds 2010; Hellbourg et al. 2012) is another interference treatment technique commonly used in pulsar search. It removes near-field RFI by using their spatial signatures in multiple beams, antennas, or receivers. There are also methods like Zero-Dispersion-Measure (zero-DM) wideband impulse clipping (Eatough et al. 2009; Keane et al. 2010) and birdy list zapping (birdy is a name for interference that occurs periodically). The zero-DM technique is commonly used in searching for transient signals such as pulsar single pulses and Fast Radio Bursts (FRBs). Birdy zapping is used for the post-process removal of pulsar candidates caused by periodic RFI. Recently, advanced methods have been presented for RFI detection, including techniques like SumThreshold and Singular Value Decomposition (SVD; Offringa et al. 2010) and machine learning (Akeret et al. 2017; Czech, Mishra & Inggs 2017, 2018a, b; Wang et al. 2019). There are also some studies that develop methods for certain kinds of interference. In Maan, van Leeuwen & Vohl (2020), the impulsive RFI is excised by a threshold-based method and the periodic RFI is mitigated by an inverse Discrete Fourier Transform. Also, special hardware and software have been developed to eliminate some specific kinds of RFI (Błaszkiewicz et al. 2021).

[1] http://www.cv.nrao.edu/~sransom/presto/

Most aforementioned methods focus on detecting and removing the RFI in the data instead of finding their sources. In this paper, we present an analysis procedure to study the taxonomy of RFI, to isolate and track different types of interference, and potentially pin down their origins.

We classify RFI into three main categories: (1) impulsive; (2) periodic; (3) non-stationary. In the frequency domain, RFI has narrow- and wideband types (Ford & Buch 2014). In the time domain, there are impulsive, periodic, and other randomly varying types. Impulsive RFI is a strong short pulse that can be either narrow or wideband signals (Fridman & Baan 2001); periodic RFI is cyclical, and often narrowband signals. We also describe a type of RFI that is not apparent impulsive or periodic but has non-stationary time fluctuations. In most cases, non-stationary RFI displays a special shape of Fourier power spectrum (often resembles a red noise or a quasi-periodic signal; Toumpakaris et al. 2003; Szadkowski & Głas 2016).

Our method is based on component analysis and temporal (Fourier) analysis. First, it isolates RFI of different origins by decomposing the original data. Secondly, it categorizes them based on the extracted RFI features, and finally, it records them in a catalogue for further statistical and spatial analysis.

RFI taxonomy and isolation could help researchers track down the sources of interference. Impulsive interference often comes from electrical discharges; narrowband or periodic RFI at some specific frequency is often caused by radio service and communication transmitters (Ford & Buch 2014; An et al. 2017); satellites can produce strong signals in a certain frequency range. By studying RFI taxonomy, categorizing, and tracking individual RFI, engineers could take measures to pin down the RFI sources by correlating the RFI with other environmental information. A good example of this is the FRB-alike millisecond-duration RFI called 'Perytons' detected by Parkes (Burke-Spolaor et al. 2010; Kocz et al. 2012; Kulkarni et al. 2014). They were identified as RFI through their multibeam signature, but their causes had puzzled astronomers for years. Eventually, astronomers found out that they came from a microwave oven by matching their occurrence time and human activity (Petroff et al. 2015).

The paper is arranged as follows: the method of component and temporal analysis is introduced in Section 2. Our experiments and the results are presented in Section 3. Finally, we discuss the functionality, caveats, and future works of our methods in Section 4. We discuss improvements made to enhance the sensitivity of the classifiers in Appendix A.

## 2 METHOD

The radio data used in this paper are composed of many quickly sampled spectra. To extract the intrinsic spectral features of different RFI, we employ a commonly used component analysis method – the Principal Component Analysis (PCA). PCA could decompose the spectral data into a set of independently varying bases. Since RFI are often the strongest varying signals in data, the most prominent bases could represent the spectral shape of the different RFI components. The varying weights of these bases represent the RFI signal's temporal variations. We can then categorize the RFI into different types by using their temporal and spectral features, and then summarize them by time and telescope pointing directions to search for spatial and temporal correlations. We combine this component, temporal, and spatial analyses to form a more complete picture of the telescope RFI environment and look for the origin of the prominent RFI. Fig. 1 shows the schematics of our RFI classifier (a specific process flow diagram is shown in Fig. A6).

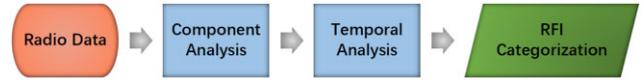

**Figure 1.** The general data processing procedure of the classifier.

### 2.1 RFI classification methods

#### 2.1.1 Component analysis

PCA is an unsupervised machine-learning method for data decomposition or compression (Bryant & Yarnold 1995). It has been shown that one could use PCA to decompose radio astronomy data into independent components, and to remove RFI through identifying and zeroing out the RFI-contaminated bases (Zhao, Zou & Weng 2013; Nguyen & Tran 2016). Here, we use PCA to identify RFI-contaminated signals, then record and study these RFI's features.

We use PCA[2] to extract principal components. If taking the radio data as a 2D (temporal and frequency) matrix $A_{m \times n}$, then we can use the SVD method to decompose the matrix into the product of three matrices (Shlens 2014):

$$A_{m \times n} = U \Sigma V^T \quad ,$$

where $U$ is the so-called *left singular vectors*, $\Sigma$ is the *singular value matrix*, and $V$ is the *right singular vectors*. In linear algebra, $V$ is the eigenvectors of the covariance matrix $C_{n \times n}$ which can be derived through

$$C_{n \times n} = A^T A$$
$$= (U \Sigma V)^T U \Sigma V = V \Sigma^2 V^T.$$

Calculating the eigenvectors of the covariance matrix is the process of diagonalizing the matrix. Therefore, the obtained eigenvalue is the diagonal element of the covariance matrix, that is, $\sigma_j$ in $\Sigma$ is the variance of each measurement dimension. As the eigenvectors of $C_{n \times n}$, $V = \{\hat{v}_1, \hat{v}_2, ..., \hat{v}_i, ...\}$ is a series of orthogonal basis. Also, these bases are the principal components extracted from $A$. For radio data, these prominent bases are often dominated by RFI.

$A_{m \times n}$ is set of a 2D radio data containing $m$ time samples and $n$ spectral channels. $A_{m \times n}$'s PCA eigenvectors $\hat{v}_i$($n$ elements) defines a set of spectral bases. These spectral bases represent linearly independent components. Since we decompose the data along the time axis, the resulting spectral vectors tend to vary independently in time.

Man-made RFI often affect a single or a small range of channels and vary significantly in time. They tend to show up in the PCA basis as out-lying points in certain channels. Thus, we can identify these narrowband varying RFI from the outliers of the PCA basis $\hat{v}_i$. In our experiment, the threshold for RFI flagging in $\hat{v}_i$ is $3\sigma$ (over the mean values by 3 times standard deviation).

For simplicity, we often rank $V$ by the significance of the diagonal terms of $\Sigma$, and keep only the most significant $k$ basis. In this paper, we choose $k$ dynamically such that $\sum_{j=0}^{k} \sigma_j^2 > 0.9$.

---

[2]*sklearn* python library (Pedregosa et al. 2011).

### 2.1.2 Temporal analysis

We first identify impulsive RFI. As a projection of the basis $v_i$'s dominated RFI on time-domain, $a_i$ is the representation of the temporal RFI. To identify impulsive RFI, we find any $a_i$ above a threshold of $9\sigma$ to detect impulsive outlier. In Section 3.3.2, we have more explanation about the threshold choosing.

To identify periodic RFI, we then check the FFT of $a_i$ for any significant peaks. Significant peaks in the Fourier transformed $a_i$ indicate the presence of a periodic or quasi-periodic RFI. Sometimes, red noise could lead to significant peak towards the lower frequency portion of the Fourier spectrum. We design a combined threshold to distinguish periodic RFI from red noise more accurately. This combined threshold is derived by considering both the outliers and the dispersion of their distribution. A detailed introduction about the combined threshold is presented in Section A3.

Finally, there is a third type of RFI that is neither impulsive nor periodic, this type of RFI shows stochastic temporal evolution that is non-stationary. Therefore, we call this type non-stationary RFI. In our practice, we categorize all RFI with no significant outliers in $a_i$ as the non-stationary type.

With the above analysis, we could study the RFI taxonomy. We develop a classier to categorize the RFI into three aforementioned types and then record them into an RFI data base based on above component and temporal analysis.

RFI taxonomy would be useful for the forensics of RFI sources. We could trace or identify the potential RFI sources through their temporal signature or their emitting frequencies, or their occurring time and directions.

For now, we categorize our detected outliers into three different general categories: impulsive RFI, periodic RFI, and non-stationary RFI. Within the categories, we also identified sub-categories such as wideband impulse and narrowband impulse. As pulsar astronomers, we are mostly interested in and affected by quickly varying time-domain RFI, this is the reason that we choose to categorize RFI first based on their temporal signature rather than on their narrow-bandness. Our analysis and taxonomy strategies are based on these preferences. For astronomers in other fields of research, they may be more interested in understanding other types of RFI that are not well characterized in our approach, e.g. stationary narrowband RFI. For their study, a different categorization may be preferred.

## 2.2 Classifier sensitivity

In order to estimate the sensitivity of our classifier, we simulate RFI signals and insert them to a white noise background. Then we test at what relative strength RFI methods can pick out.

In our simulation, we generate a 2D mock data set with 32 768 samples in time and 4096 channels in frequency, following the same shape as the FAST 19-beam pulsar search data. The mock data set are filled with a background noise of normal distribution and a standard deviation of 1 and a mean of 0. We then inject three kinds of RFI in our mock data: narrowband non-periodic RFI (we call it narrowband RFI in this section) and periodic RFI in a single channel, and wideband pulse in the whole band. The strength of injected RFI varies in a wide range. We use the unit of dB, which is defined by the ratio of the injected signals and the noise, to measure the strength of injected RFI. Here, dB is measured through $10 \, log_{10} \, (P_{\rm rfi}/P_{\rm noise})$. We also convert dB to unit of $\sigma$ when presenting them in figures and tables for ease of understanding. We conduct tests for narrowband, periodic and wideband RFI in three different experiments. We chose the number of RFI to inject in the trial data set based on the real average event rate of these RFI types. We sampled 100 raw files, and counted how many real narrowband RFI were present in those files. The prominent narrowband RFI channel rate of the data in the FAST early stage is about 11 per cent (~440 channels). So we inject additional signals into 400 channels at each trial when simulating narrowband and periodic RFI independently. Considering that wideband impulsive RFI is not frequently happening, we only inject 1 wideband pulse per data set. Every type of RFI has 1000 trials.

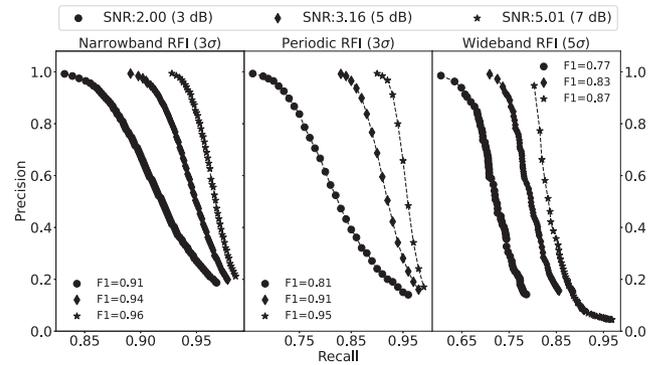

**Figure 2.** The precision-recall curves for the three types of simulated RFI: left-hand panel: Narrowband RFI; middle panel: Periodic RFI; right-hand panel: Wideband RFI. For classifying narrowband and periodic RFI, recall and precision of the classifier are sampled from 400 channelized RFI and the classifier threshold is chosen to be $3\sigma$. For classifying wideband (impulsive) RFI, there is only 1 pulse in each data file and the classifier threshold is set to $5\sigma$. These thresholds were chosen based on the theoretical noise floor of the simulated data (see text for details). We also label the best F1 value of the curves as well as the SNR of injected RFI.

It is only sensible for any classifiers to detect RFI signals above the noise floor of the data, therefore our classifier threshold is chosen as such in these mock experiments. Here, we define noise floor as the threshold level above which there is one false positive event expected from the given number of trialing time samples or frequency channels. For narrowband RFI, the noise floor is determined by the number of channels (4096), e.g. the noise threshold for having one false positive event in 4096 trials is $3.67\sigma$. For wideband RFI, the noise floor is determined by the number of time samples in a data set, and is about $4.17\sigma$. Based on the above estimation and our practical experience, we choose the following signal thresholds for our classifier when dealing with the narrowband ($3\sigma$) and wideband RFI ($5\sigma$).

Simulated RFI is injected into the 1000 mock data sets with the same strength, e.g. $2.00\sigma$ (3 dB), $3.16\sigma$ (5 dB), $5.01\sigma$ (7 dB). Then we measure the recall and precision of the PCA classifier. Recall is the ratio of the detected RFI to the total number of RFI channels or pulses. Precision is the ratio of the flagged RFI to the total detected RFI numbers. Then we measure the recall of the detection of these RFI types from the 1000 trials.

The statistics of recall and precision are presented in Fig. 2. Three panels presented the precision–recall curve for detecting narrowband, periodic and wideband RFI, respectively. For the simulated narrowband and periodic RFI, we use our PCA classifier with a signal threshold of $3\sigma$, while for the wideband RFI we choose a threshold of $5\sigma$ based on the aforementioned noise floor analysis. For each panel, we test three different levels of noise injection, the details are shown in the figure. For each chosen RFI signal strength, we calculate the maximum F1 score along the precision-recall curve and labelled them on the plot.

**Table 1.** RFI information recorded in the RFI data base. All the RFI have been recorded together with their occurrence-time, frequency, and spatial information. Additionally, the RFI data base records the S/N of pulses for the impulsive RFI and the period of the periodic RFI.

| | Time info | Frequency info | Spatial info | Others |
|---|---|---|---|---|
| Impulsive RFI | UTC of pulses | | | S/N of pulses |
| Periodic RFI | UTC of observation | RFI channels | RA, Dec. and Alt, Az | Period of RFI |
| Non-stationary RFI | UTC of observation | | | – |

**Table 2.** Data information of the three experiments. All these data are taken by pulsar backend with a sample time around ∼200 μs.

| Experiments | Date | Receiver | Frequency (GHz) |
|---|---|---|---|
| **One day RFI-identification** | 2019.05.23 | 19-beam | 1.02-1.48 |
| **RFI masking for pulsar search** | – | UWB/19-beam | 0.2-1 (UWB) |
| | | | 1.05-1.45 (19-beam) |
| **Commissioning drifting scan** | 2017.09-2018.04 | UWB | 0.2-1 |

The simulation result shows that PCA is more effective to detect narrowband RFI than wideband RFI.

## 2.3 Statistical and spatial analysis

We record the meta information for each type of RFI along with their classifications. For example, we record the receiver orientation of the data to analyse the RFI's spatial information. We mark the accurate time of arrival for RFI to obtain their possible temporal evolution. Table 1 lists the specific information of the three RFI types that we store in the outputs of the classifier.

We also use the meta information of the RFI in the data base to capture their date-time and spatial distribution. Then we correlate them with possible RFI signal sources and human activity cycles. We compute the multidimensional distribution of the categorized RFI over the occurrence time, frequencies (i.e. channels), and the RFI components' variance ratio, which indicates how serious the RFI is in comparison with the rest of the PCA components. We study the S/N distribution of impulsive RFI, the period distribution of periodic RFI, and analyse them alongside the RFI date-time distributions. In doing so, we could potentially identify RFI related to human activity. By analysing the RFI spatial distribution, we could pinpoint the directions of the RFI sources. To prove these concepts, we applied this method to drift scan data taken by FAST in the commissioning phase, and present the experiment results in Section 3.

## 3 RESULTS

In order to test our analysis methods on real data, we conducted three different experiments: (1) *One-day RFI-identification Experiment*, (2) *RFI Masking for Pulsar Search*, (3) *Commissioning Drifting Scan Experiment*. In *One-day RFI-identification Experiment*, we tested PCA and temporal analysis; in *RFI Masking for Pulsar Search*, we tested PCA bad channel identification; in the *Commissioning Drifting Scan Experiment*, we tested RFI taxonomy and spatial analysis. Table 2 lists the set-up of the observations used in our experiments. All our testing data were taken in PSRFITS format (Hotan, Van Straten & Manchester 2004) with 8-bit sampling. During the commissioning period, FAST uses two different receivers to collect signals, which are ultra-wideband receiver (UWB) and a 19-beam receiver.

**Table 3.** A planned timetable of the RFI experiment, with some devices omitted due to the list being too long to list here. The exact power-on time might deviate for a few minutes from the plan. Specific functions of these devices have been introduced in Nan et al. (2011), Wang et al. (2018), and Zhang et al. (2018).

| Device | Power-off time | Power-on time |
|---|---|---|
| Total station | 09:50 | 16:25 |
| Actuator | 10:43 | 16:06 |
| Monitoring equipment | 11:00 | 16:25 |
| Relay rooms | 12:30 | 16:05–16:52 |
| Electricity in workshops | ∼12:52 | 16:06 |
| ⋮ | ⋮ | ⋮ |
| Power equipment for towers | 14:55 | 16:05–16:52 |
| 1∼ 6#Capstans | 15:35 | 16:05–16:52 |

### 3.1 Experiment i: one-day RFI-identification experiment

The FAST radio environment monitoring team conducted a one-day RFI identification experiment on 2019 January 3. At the beginning of this experiment, they powered off most electronic devices in the vicinity of the telescope. Then they turned these devices on one by one at specific times during the day to make sure which ones were causing RFI. Table 3 lists the turn-on time for some devices.

We use the data taken in this experiment to test our PCA and temporal analysis scheme. By identifying the onset time and spectral feature of some prominent RFI, we try to cross-reference them with the device power-on-off timetable. Fig. 3 shows the results of our component analysis. For the top four components, the variance ratio is 57.4 per cent, 18.8 per cent, 3.2 per cent, 2.3 per cent in descending order. The first component is composed of the persistent signals including the bandpass of the data. The next three components may be caused by individual varying sources. Fig. 3 shows that the weight of component 1 gets peaks at 15:35 and 16:13, and the basis of this component has peaks at frequency 1056, 1088, 1120, 1420, 1421 MHz. The basis of component 2 shows spikes at 1056, 1088, 1120 MHz. What corresponds to these spikes on the basis is the weight's turning-off at 9:50 and turning-on at 16:25. This temporal behaviour matches the on-off time of the total station, indicating that the component is likely caused by the total stations. Interestingly, component 3 shows spectral spikes around 1420, 1421 MHz and slowly peaked at around 16:00, this is caused by the H I emission from the Galactic plane. Similarly, the weight of component 4 (1485–

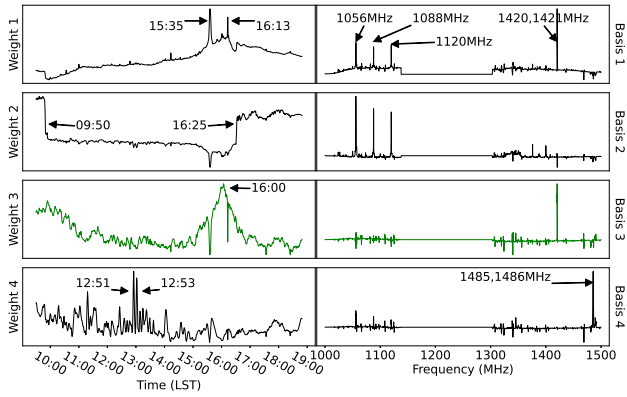

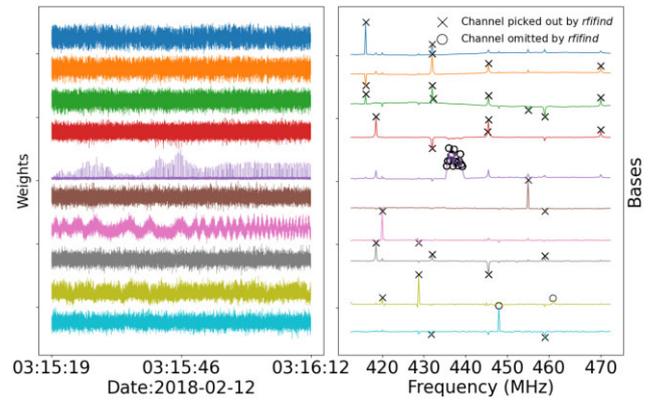

**Figure 3.** PCA analysis result of the top four components. Channels between 1137 and 1303 MHz were excised in the data processing because of significant contamination from satellite signals. The left four panels show the varying weights of the components, the right four panels show the spectrum bases of the components. The sources of component 2 and 4 have been identified by cross-referencing with Table 3, while component 3 is caused by the 21 cm line of neutral hydrogen H I, so we plot component 3 with a green line to distinguish the astronomical component with other RFI components.

**Figure 5.** Result of a component analysis on data. Only the top 10 weights and bases are represented. In the right-hand panel, we marked the channels that are over $3\sigma$. The channels marked with circles are the RFI marked by PCA but missed by the *rfifind*.

Fig. 4 displays the RFI (and H I) identified by PCA in the format of the raw data.

Based on the aforementioned PCA and temporal analysis, we concluded that:

(i) The switching of power of the capstans might cause broad-band transient interference (as seen in 15:35 and 16:13 h).

(ii) The FAST Total Stations produced interference at frequency 1056, 1088, 1120 MHz.

(iii) The component analysis also identified the 21 cm line of neutral hydrogen H I at frequency 1420 MHz from the Galactic plane (passing meridian around 16:00 h).

### 3.2 Experiment ii: RFI masking for pulsar search

RFI masking is a critical step for most pulsar search projects. For example, the *rfifind* program in PRESTO detects and masks RFI by using a moving-statistics threshold filter. However, some RFI could still pass the threshold filter in *rfifind* and generate false candidates or mask real signals.

Our PCA method could be used to identify spectral channels contaminated by weak RFI, especially by repeating or periodic ones. In this subsection, we use pulsar search data collected by FAST to test the RFI masking function of our method. We applied the PCA to the test data and marked bad channels for zapping with RFI bases (i.e. the basis of the decomposed component) crossing a threshold of $3\sigma$.

In this experiment, we compare the RFI channel masking result between *rfifind* and PCA. Fig. 5 shows the PCA analysis of our test pulsar data. In this figure, we present the top 10 components. The channels marked as circles or crosses are RFI channels over the $3\sigma$ threshold on the basis. Though *rfifind* detected most of the RFI channels, a periodic interference around 437 MHz in the 5th component was not marked by *rfifind*.

We searched the data using PRESTO with only *rfifind* as RFI excision method, and a pulsar candidate caused by the periodic RFI was found and shown in panel (a) of Fig. 6. Then we folded the data with the same period after zapping the bad channels identified with PCA. The result is shown in panel (b) of Fig. 6 where the periodic RFI around 437 MHz now disappeared.

Beyond masking data, *rfifind* could also output a list of 'birdies' to help remove pulsar candidates caused by periodic RFI signals. The birdy list removes all candidates in the specific rotation period range,

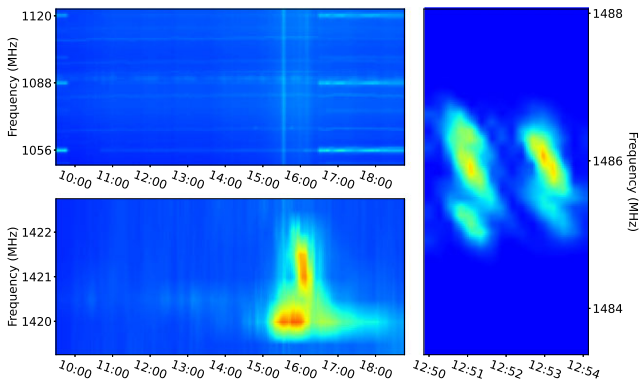

**Figure 4.** A zoom-in view of the raw data. Top left panel: component 2, strong peaks around 1056 MHz, 1088 MHz,1120 MHz disappeared at 09:50 and came back again at 16:25. In this panel, we could also see broad-band transient RFI at 15:35 and 16:13. Bottom left panel: i.e. H I signals around 1420,1421 MHz came in roughly at 16:00 and lasted for tens of minutes. Right-hand panel: component 4, i.e. interference at 1486 MHz spiked around 12:51 and 12:53.

1486 MHz) spikes at 12:51 and 12:53, which seems to match the onset time of workshop electronics.

From the above PCA analysis, we find that component 1 seems to contain signals from both the total station and the H I emission, indicating the decomposition of the signal by PCA is not entirely clear. The basis of component 1 might have some signals from components 2 and 3. However, the peaks around time 15:35 and 16:13 that we see in the weight of component 1 do not show up in the weight curve of components 2 and 3, indicating that this might be caused by some other reasons. Upon a closer inspection of the raw data (see top panel of Fig. 4), we find that there were two broad-band transient signals around 15:35 and 16:13 h. Therefore, the peaks in the weight of component 1 are likely caused by broad-band signals generated around this period. Cross-referencing with the planned timetable, we found that component 1 might be linked to the power-off and power-on of the capstans that pull the receiver cabin (Nan et al. 2011).

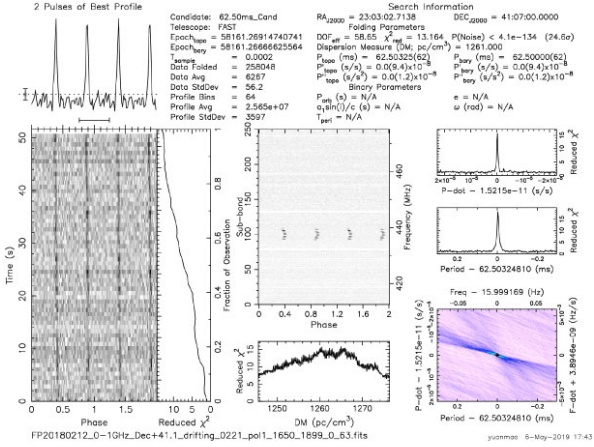

(a) *rfifind*.

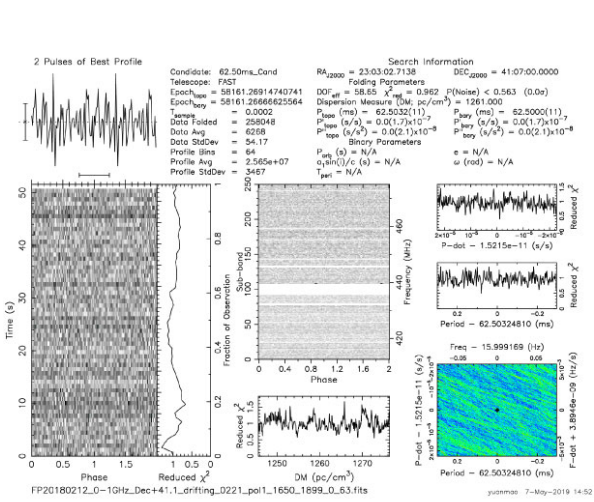

(b) *rfifind* & PCA.

**Figure 6.** The pulsar search testing for two RFI masking techniques. The top panel (a) is a result of masking RFI only by single *rfifind*. It shows that the undetected RFI causes a negative pulsar-like candidate. Bottom panel (b) shows the search result with a combined usage of *rfifind* and PCA. It shows that the pulsar-like fake signal is eliminated from the candidates.

usually after these candidates were found and folded. Our PCA-based method could remove the RFI before searching the data and only removes the data in the narrowband affected by the RFI. As this experiment shows, the component analysis can detect the periodic signal effectively even though the signal power is not strong.

Periodic RFI, i.e. 'birdies' are common and sometimes non-trivial to excise because they generally do not cause significant spikes in data, and are often only visible when analysing high-time-resolution data, such as pulsar search data. In our PCA analysis of the pulsar search data, we identified a number of such periodic/quasi-periodic RFI. Fig. 7 shows fake pulsar candidates caused by birdies that are recorded by RFI classifier. These two fake pulsar-like signals are caused respectively by periodic RFI (panel a) and a quasi-periodic (the period drifts with time) RFI (panel b). More details of these two detected RFI are described in Section 3.3.3.

By zapping the bad channels associated with these birdies or ignoring these recorded birdies, the pulsar research program will no longer cost time and computing resources to search and fold these

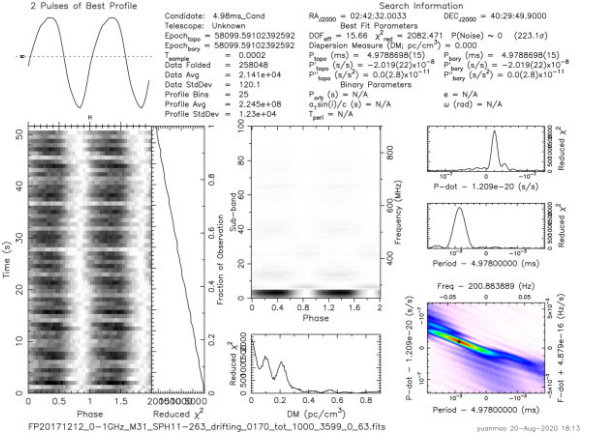

(a) candidate caused by period-stable birdy

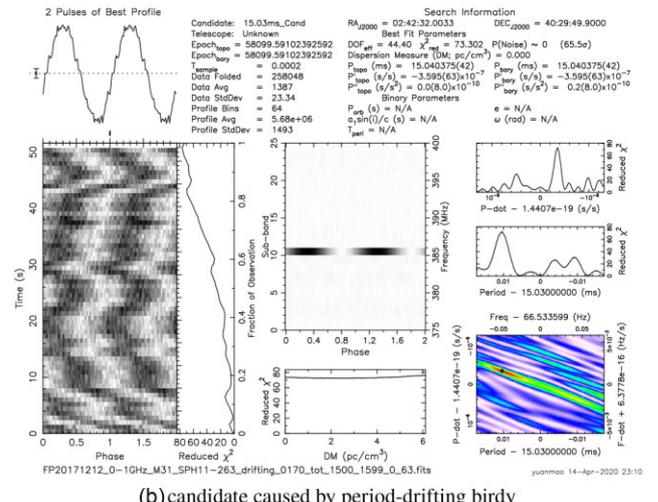

(b) candidate caused by period-drifting birdy

**Figure 7.** Fake pulsar candidates caused by the period-stable and period-drifting RFI.

periods. This method has been applied to the pulsar search pipeline for the CRAFTS survey.

### 3.3 Experiment iii: commissioning drift scan

This experiment is a long-term process of RFI detection, classification, and analysis from FAST's commissioning drift scan data. The purpose is to study the radio environment around FAST and test the practicability of our classifier. The FAST commissioning team conducted a pilot pulsar survey program at night (roughly 8 pm–8 am) during 2017 September to 2018 April by operating the telescope and recording data from FAST-UWB receiver in a drift scan mode (Di Li et al. 2018). In this case, the telescope's azimuth is fixed and only has daily elevation changes: the north (with $Alt \sim 360°$ or $0°$) or south (with $Alt \sim 180°$) direction. Nevertheless, these data enable us to statistically study the rise and fall of RFI background from dusk to dawn.

In this experiment, we first introduce the performance of the classifier in Section 3.3.1. After the description of the RFI classifier, we estimate the classification precision of the classifier in following Section 3.3.2. At last Section 3.3.3, we present the spatial analysis of some RFI categorized by the classifier. In section of Appendix A, we present some improvements of the component analysis we made to this classifier, for more accurate classification. These improvements include additional discrimination and selection to narrowband RFI, wideband pulses, and periodic RFI.

### 3.3.1 The real-data performance of the RFI classifier

In Section 2.1.1, we introduced the concept of the RFI classifier. We combine PCA component analysis and time-domain Fourier analysis to classify and categorize RFI. According to the time-domain characteristics, the classifier categorizes most RFI into one of three categories: (1) Impulsive, (2) Periodic, (3) Non-stationary, and record their information into an RFI catalogue. Specific information items are listed in Table 1.

In order to visualize these different types of RFI, we designed a suite of graphical diagrams for summarizing the RFI components in a given data set. Fig. 8 (including sub-figures a, b, c, and d) is an example of such graphics based on a set of pulsar data taken on 2018 February 22.

This classifier produces four illustrations: Sub-figure (a) displays a summary of the data set. It exhibits the most apparent RFI components. On the top left sits a two-column text table that shows the metadata such as coordinate and time (UTC) of the observation and statistical information about the RFI, such as the percentage of *contaminated channels*. On the top right are two charts, the pie chart shows the partition of RFI by their variance. The curve chart shows the bandpass of the data with detected RFI channels masked as the red cross ($3\sigma$-crossing points in the component bases). The bottom part contains three displays. On the left are the varying weights of the components, on the right are the component bases, and in the middle are the variance ratios of the components. We extract the significant components up to a total of 90 per cent variance and keep the RFI information of the component whose variance ratio is greater than 1 per cent. In this way, we often end up with a few hundred components to categorize and record. As a visual output, we only present the components whose variance ratio is over 1 per cent in this panel.

Sub-figures (b), (c), (d) display the impulsive, periodic, and non-stationary types of RFI, respectively. In all these panels, we show only the top 10 or less components. In each sub-figure, we put meta information, including the number of components and total variance, on the top left corner, then two line charts on the top middle and the top right. The top-mid chart shows where the RFI emerges in time and frequency, the top-right chart show where the RFI ranks in the variance ratio. For the impulsive and non-stationary RFI, the bottom part of the panel is composed of two plots that show the varying weight and the base of each displayed component. For the periodic RFI, there are three plots in the bottom, with the additional middle one showing the Fourier spectrum of the weight curve, which contains information about the periodicity of the RFI.

### 3.3.2 The precision of the RFI classifier

In Section 2, we tested and demonstrated our classifier's sensitivity through mock data. It is a good practice to also test its accuracy on real observational data.

Here, we test our analysis method on a set of real data taken by FAST during its commissioning phase. We selected over hundreds of real data sets, containing various types of RFI. We picked subsets of the real data for the test of different type of RFI, because some of them only contain one type of RFI and not others. To get the recall and precision of our PCA classifier, we check the RFI by eye and labelled these test data by hand. For periodic RFI, we take an FFT to each channel and then check their spectrum to identify periodic signals. For impulsive RFI, we check their single-channel and integrated temporal series, to verify whether they have pulses or not. The threshold we choose for periodic or non-stationary RFI is $3\sigma$ while $9\sigma$ for impulsive RFI. We chose a much higher threshold for impulsive RFI based on two reasons. The main reason is that we get 4096 samplings for narrowband RFI, but get more samplings for wideband RFI and noise. Another reason is that the impulsive RFI is time-domain RFI, and we pick the pulses at the temporal weights that usually have a varying baseline. Therefore, we set a higher threshold on the PCA weights that we can filter the pulse candidates caused by a varying baseline. We calculated the recall and precision of our classifier on different types of interference. The results are shown in Table 4.

We found that using all components with variance-ratio greater than $10^{-5}$, generally accounts for 2/3 of the total 4096 components. A large number of components can always lead to a higher recall, but return a relatively lower precision, because the indistinctive components contain both weak RFI and noise. And in some cases, it is hard to distinguish weak RFI (e.g. quasi-periodic RFI) from noise, even by eye. Therefore, we focus on picking out the most strongly varying (or serious) RFI and categorize them correctly. We generally analyse only the components of prominent RFI whose variance-ratio is greater than $10^{-3}$ in our program, and this will also require less computing to accomplish.

In order to measure the precision of classifying the most apparent RFI in our data, we collect more than 400 wideband impulses and more than 8000 narrowband impulses from the data base, and then check the raw data file by eye to verify them. We inspected the RFI from part of the raw data used in our experiment by eye instead of RFI candidates from classifier.

For wideband impulses, we plot the data corresponding to the detected RFI location as 2D images of time and frequency, and bin frequency into eight subbands, downsample the time dimension by a factor of 16 to enhance the signals in each pixel, and checked the pulses by eye.

For narrowband impulses, we plot the data from the specific channel and time where the RFI is detected and check for the existence of RFI pulses. From our test sample, our classifier scores a precision of $94 \pm 4$ per cent for wideband impulsive RFI and $92 \pm 5$ per cent for narrowband impulsive RFI.

For periodic RFI classification, we collect more than 100 data files (every 52.4 s in length) and check the Fourier spectrum of each channel marked by the classifier for each file, respectively, and see if a genuine periodic signal exists where the classifier has suggested. One file has dozens to hundreds of periodic components after PCA decomposition, we check more than 5000 periodic RFI signals in total and found a precision of $82 \pm 13$ per cent.

It should be noted that the above F1 scores and the precision could be biased because of incomplete identifications to some weak or quasi-periodic RFI even when we check the raw data by eye. We double-check part of the eye-flagged RFI and found that there are only a few cases of such RFI in our data base (less than 3 per cent on average) that might cause misclassification. Therefore, our statistical result could be biased but at a relatively lower level.

### 3.3.3 Spatial Analysis of catalogued RFI

We save information of the categorized RFI in a data base (as shown in Table 2). As this information (such as the date, time,

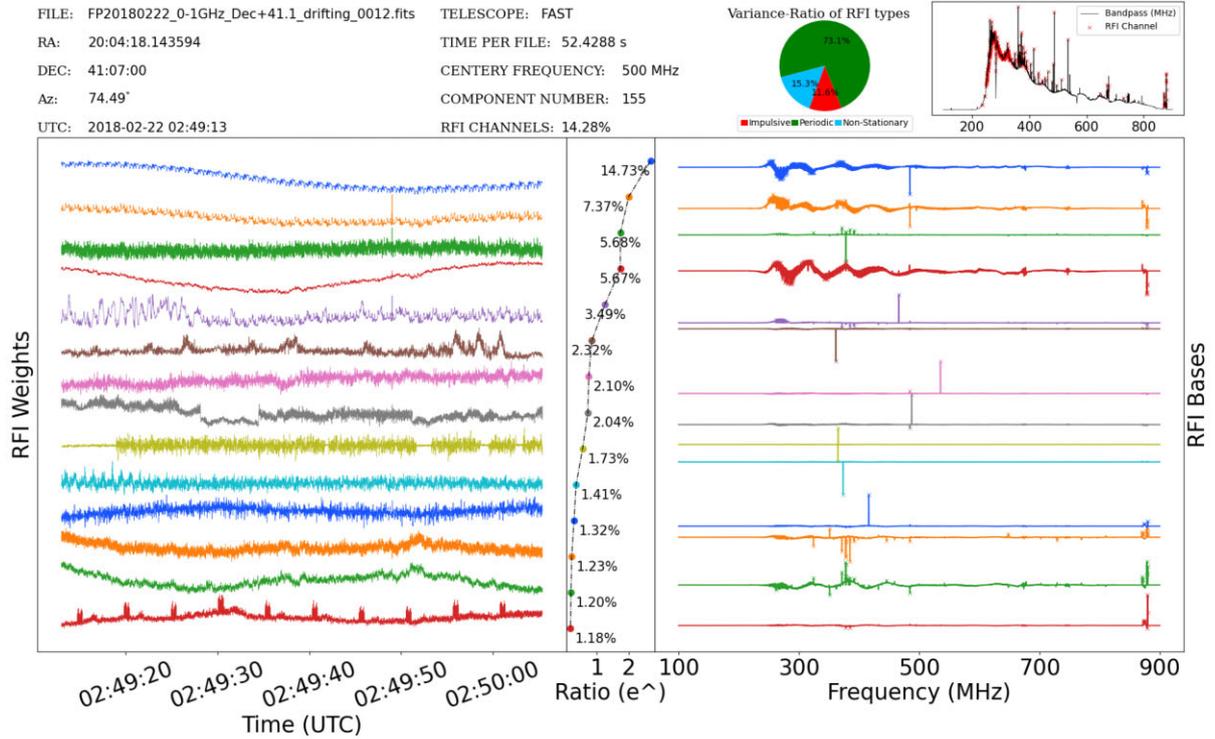

(a) Overview diagram. A summary display of the RFI situation in the data, containing both the temporal and frequency features of the RFI.

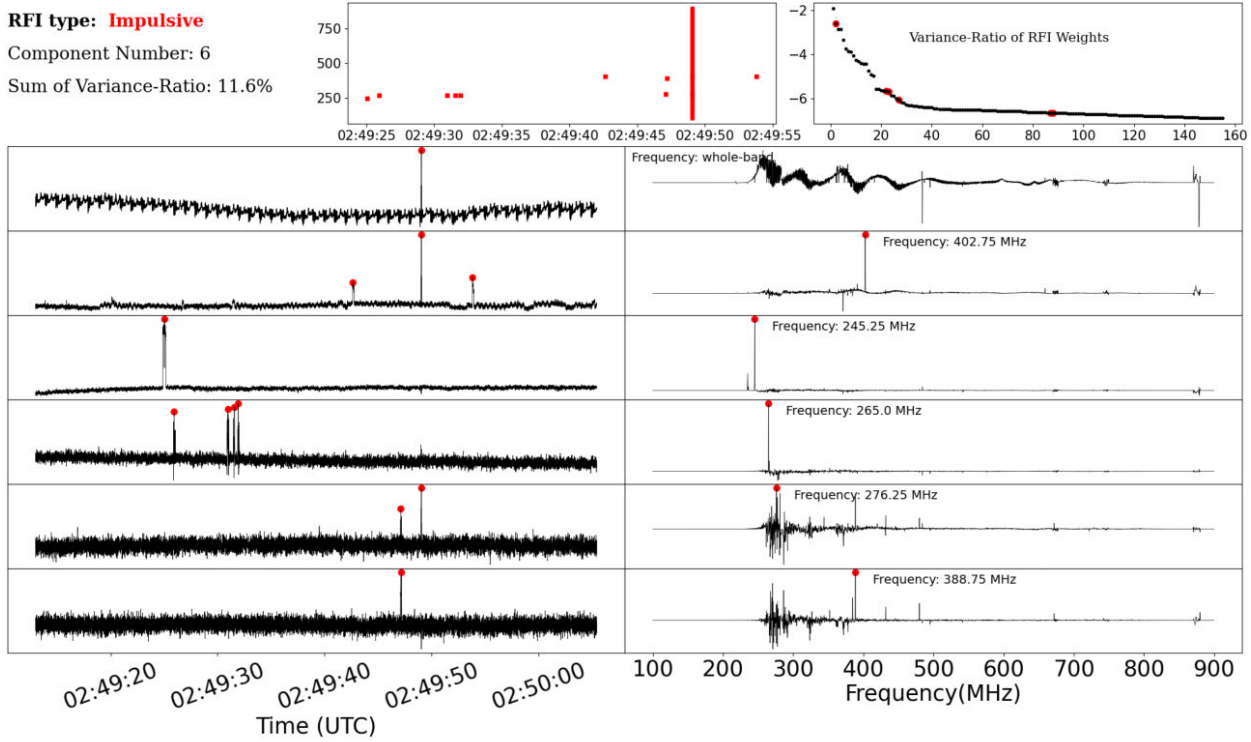

(b) Impulsive RFI.

**Figure 8.** The graphic outputs of the RFI classifier from a radio data set. There are four illustrations of the final categorization results, which are, respectively, the general RFI decomposition, the impulsive type, the periodic type, and the non-stationary type. From sub-figure (a) to (d) is the display of the four graphical results.

and telescope pointing) could provide crucial forensic evidence for some interference signals, we try to help identify the origin of these RFI. The following paragraphs are some examples of such attempts.

Temporal behaviour of impulsive RFI: Here, we show two examples of temporal behaviour of impulsive RFI that may be related to human activities.

(i) Case 1: the temporal evolution of impulsive RFI.

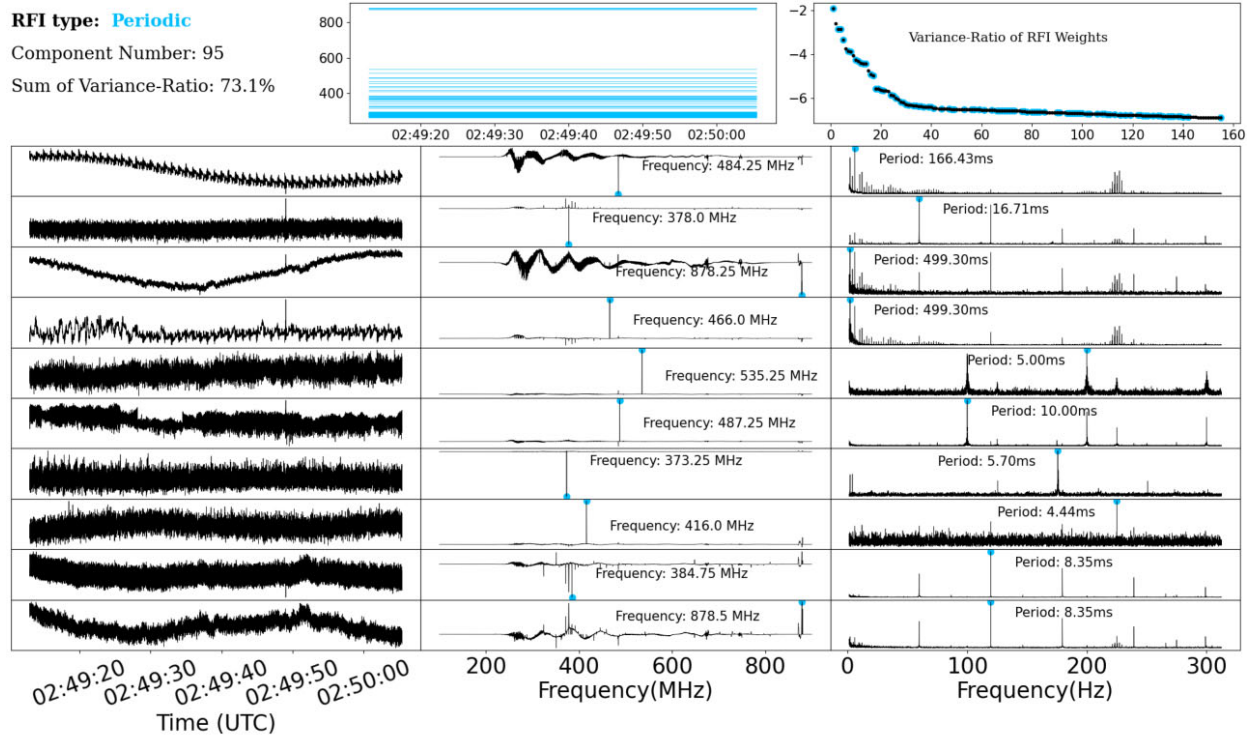

(c) Periodic RFI

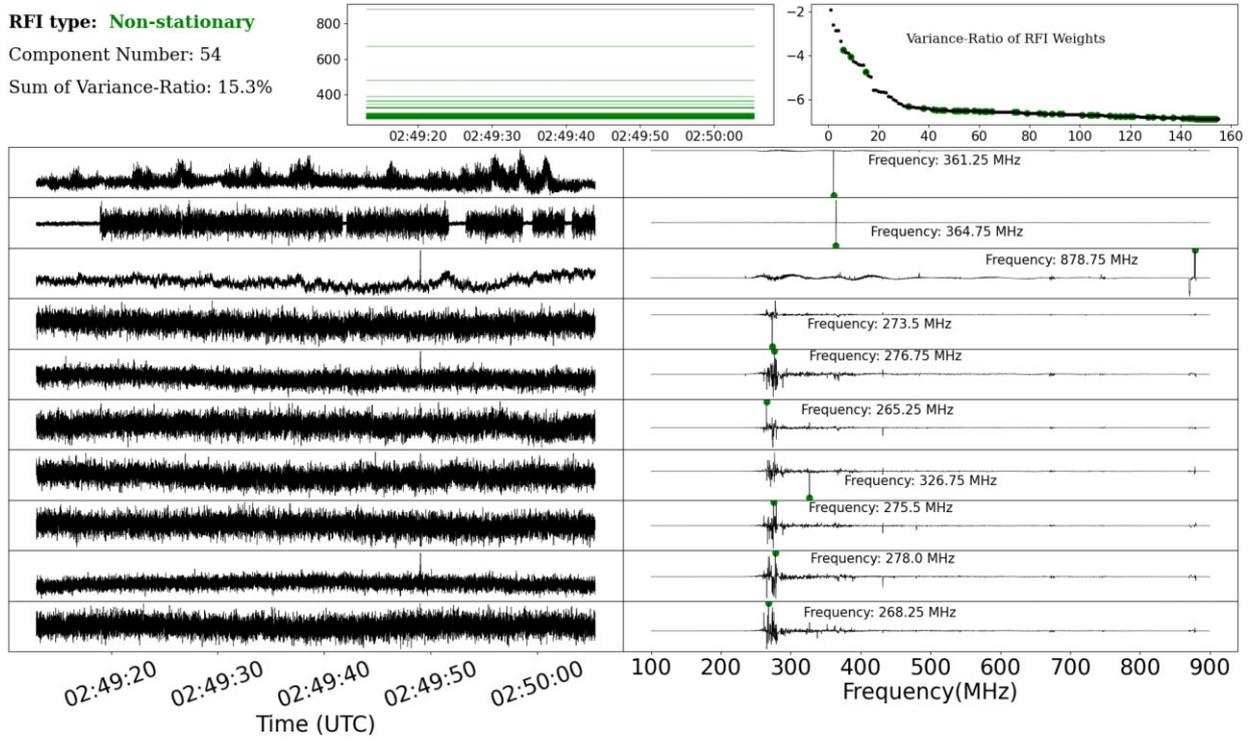

(d) Non-statinary RFI.

**Figure 8** – *continued*

In the commissioning data collected from FAST, on some days when observations were taken in the morning, we could find a rise in occurrence rate for impulsive RFI, which could be correlated with the local time and the detailed schedule of the observatory site. Fig. 9 displays such an impulsive RFI diagnostic plot for the data taken on 2017 December 7. Colours here are presentations of the variance ratio of each RFI weight, showing how significant are these RFI. In this analysis, it is apparent that around UTC 23–24 h (local time 7 am–8 the uplink CDMA communication signals. First, the uplink CDMA signal covers a frequency range from $820\sim830$ MHz; secondly, these communication signals are transient (i.e. narrowband and impulsive) signals matching the feature of the uplink CDMA signal.

**Table 4.** The statistics of recall and precision.

| RFI type | Threshold | Recall | Precision | F-score |
|---|---|---|---|---|
| Periodic | $3\sigma$ | 86.44 per cent | 64.98 per cent | 0.74 |
| Narrowband impulsive | $9\sigma$ | 74.80 per cent | 97.32 per cent | 0.85 |
| Wideband impulsive | $9\sigma$ | 85.71 per cent | 84.38 per cent | 0.85 |

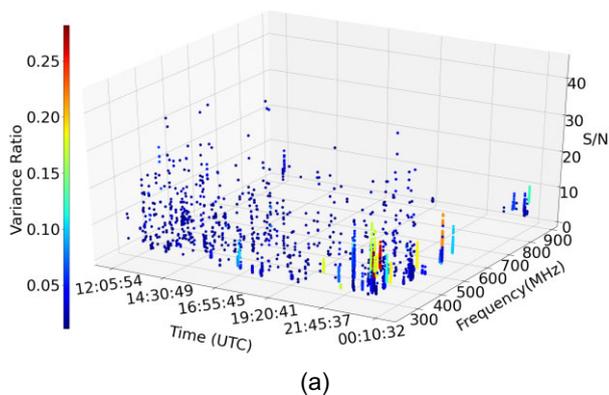

(a)

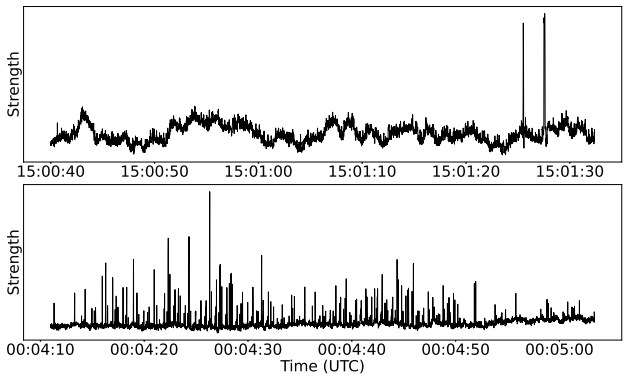

(b)

**Figure 9.** An analysis of impulsive type. Sub-figure (a) is a 3D presentation of RFI information about impulsive type recorded on 2017 December 7. Measurements on vertical 'S/N' are multiple that the pulse intensity exceeds the standard deviation of the corresponding weight sequence. Colours of these dots were decided by the variance ratio derived by PCA. Sub-figure (b) is a comparison of the severity of impulsive RFI at different times. Both the signal strength here is in an arbitrary unit. The case on the top is interference with a relatively small variance ratio. At the bottom, impulsive RFI became serious and hence have a larger variance ratio, as sub-figure (a) reveals.

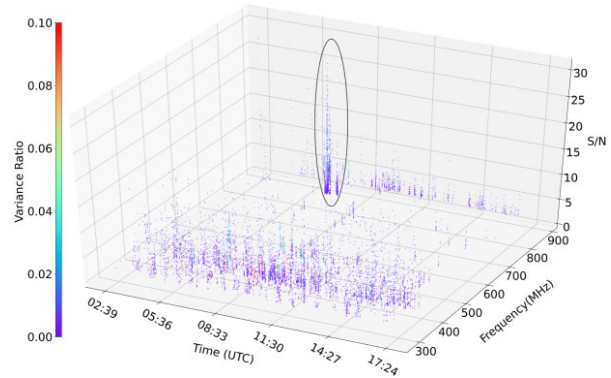

**Figure 10.** An impulsive RFI burst occurred at around 6:30 UTC (14:30 pm the local time). The frequency band of this event is around $825 \sim 830$ MHz.

Temporal behaviour of periodic RFI: Similarly, Fig. 11 shows a special example of periodic RFI diagnosis. This diagnostic figure shows how the significance, period, and channels of all the identified periodic RFI changes with time in one long observation. In this special example displayed in the figure, we see multiple drifting periodic RFI signals that appear to be varying with time in a similar manner. Their occurrence frequency changes from 60 to 100 Hz (sub-figures a and b). Different signals seem to have different radiation frequencies (detected in different narrow observational channels in 225–400 MHz range, shown in sub-figure c). We also see another set of periodic RFI that occupies stationary spectral channels and occurrence frequencies (the time-independent period shown in sub-figures a and b).

These stationary periodic RFI are more common, while the drifting ones are special. We found such period-drifting RFI in only a few observations, and their origin is still unknown. But such RFI would significantly affect pulsar searches because they cannot be easily rejected by an existing birdy list. Therefore, it is necessary for us to monitor and identify such drifting periodic RFI in our data.

To further illustrate our point, we searched segments of data containing the above periodic RFI using PRESTO and found that these periodic RFI do manifest as period candidates. Fig. 7 is the searching result of the periodic RFI. The classification method that we introduce in this paper could help us dynamically identify the existence of such RFI and their emitting spectral channel. We could remove these false pulsar candidates by masking these affected channels, vastly reducing their contamination to the pulsar searches.

Spatial distribution of impulsive RFI: Using the classification data base, we could correlate RFI occurrence with the telescope pointing to analyse the spatial distribution of RFI. However, during the commissioning phase, the FAST telescope was mostly pointing at the meridian, only changing its elevation angle. It means that the am), impulsive RFI between frequency 400–700 MHz starts to rise, possibly due to construction work during the commissioning phase. Unfortunately, the phenomenon we observed in this figure came from post-processing of historical data, which prevented us from further investigating the causes of the RFI. The method demonstrated here is to be implemented in a semi-real-time manner and will be useful for identifying emerging RFI sources. Panel (b) of Fig. 9 shows two different impulsive RFI components, which are respectively from the impulses at around 15:00 and 00:00 (UTC) in panel (a).

(ii) Case 2: Bursts of impulsive RFI.

Fig. 10 shows a cluster of impulsive RFI bursts caught on 2018 November 12. We circle these burst events with a black oval in the figure. The bursts came from around $825\sim830$ MHz, at around 14:30 h the local time. We speculate that this event was caused by observation is only pointed either at the South or North. So we can only study the difference in RFI environments at the two opposite directions.

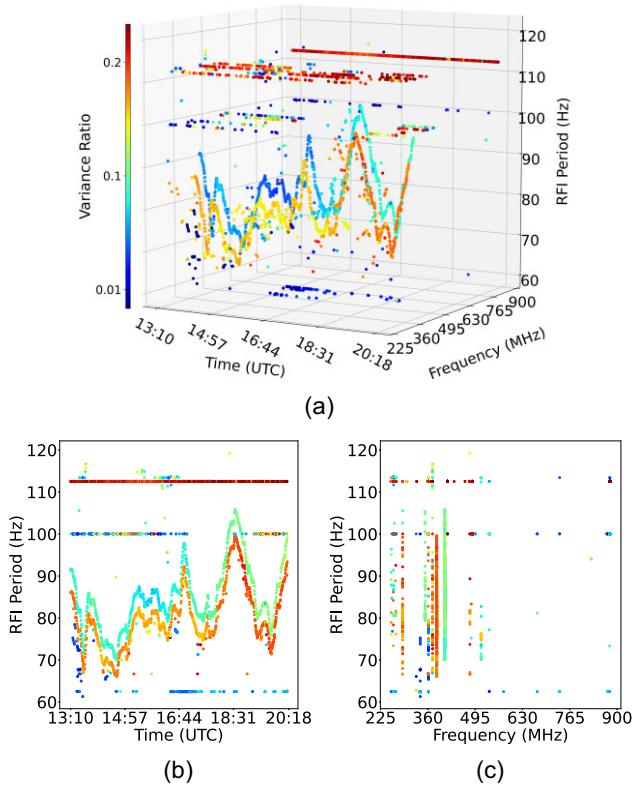

(a)

(b)                    (c)

**Figure 11.** Sub-figure (a) is a 3D representation of statistics of the periodic type. The RFI period on vertical axis is the frequency index corresponding to the maximum value of the FFT spectrum. In this figure, there are two kinds of periodic RFI. One is period-stable and the other is period-drifting. Sub-figures (b) and (c) are, respectively, the projection of 3D figure on time, and frequency band axis. These 2D plots show that the period-stable RFI (with the Fourier frequency at 100,113 Hz) happened at almost the whole band; and the period-drifting kind exited at only some specific channels, specifically at 258, 385, 512 MHz.

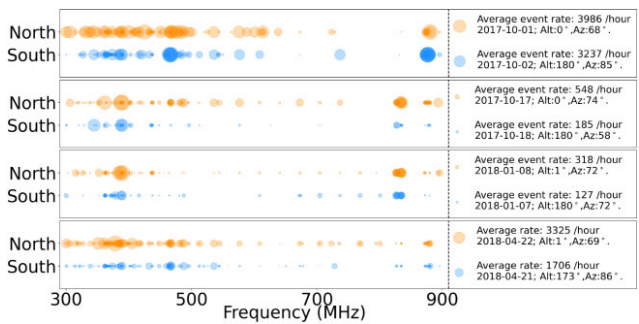

**Figure 12.** A spatial statistic about the impulsive RFI. The impulsive-RFI event rate is obtained by dividing the number of transit pulses by the total observation time (in hours). The orange and blue circles, respectively, represent the number of pulses from the North and South.

In Fig. 12, we show the event rate of impulse RFI for four pairs of observations. Each pair are two observations taken in two consecutive days but pointing in opposite directions. It indicates that there are generally more impulsive RFI from the North than from the South. Also, the RFI that came from the North covers a more diverse range of spectral channels. This may be geographically explained that the closest town (Kedu) lies to the North of FAST, and Guiyang, the capital of Guizhou Province, is also located to the North of FAST.

## 4 CONCLUSION AND DISCUSSION

In this paper, we present a set of RFI analysis methods based on the PCA and FFT. We combine these methods to develop an RFI classifier that could categorize RFI into three different general types: impulsive, periodic, non-stationary. Our method relies on principal decomposition to separate the signals of different RFI sources, allows us to detect weaker RFI than transitional threshold-based method (such as *rfifind*). Then we perform Fourier analysis to detect periodic RFI that is detrimental to pulsar searches. Finally, we classify the RFI based on the features extracted by PCA and Fourier analysis, then use these extracted features to build a more complete picture about the telescope RFI environment. By correlating the RFI analysis results with the telescope meta information, we could perform forensics to track down the RFI sources.

Our method for categorizing the RFI is customized to the pulsar search program of FAST, but also could be generalized to other telescopes or science cases.

In Section 3, we present three experiments that show the effectiveness of our methods.

In the first experiment, we use data taken in a one-day experiment dedicated to finding the prominent RFI sources around the observatory compound. Our experiment shows that PCA method is able to extract the spectral and temporal features of individual RFI, including the exact time and spectra channel in which the RFI occurred. This helped us identify at least two RFI sources.

In the second experiment, we take the PCA analysis to help mask bad channels in pulsar search data from the FAST-UWB receiver. The PCA help to build a more complete list of RFI contaminated channels, including channels with relatively weak RFI that is not detectable by the transitional threshold-based method in *PRESTO-rfifind*.

In the third experiment, we perform temporal and spatial analysis on the RFI categorized based on the classifier from the commission drift scan FAST-UWB data. We find potential connections between the rise of impulsive RFI and human activities. The result indicates that impulsive RFI is stronger when the telescope is pointing towards the general directions of nearby towns and cities.

Beyond these practical experiments, we also analyse the sensitivity of the classifier. We collected a random set of files and measured the precision of the RFI classifier. For the impulsive RFI, the precision of the classification are $92 \pm 5$ per cent for wideband pulse and $94 \pm 4$ per cent for narrowband pulse. For the periodic type, the averaged accuracy rate is $82 \pm 13$ per cent.

It should be noted that the main goal of this classifier is to quickly extract and analyse the main interference of radio data without costing too much computing resource, rather than to perform detailed detection and identification of each interference. So the classifier may cause false-detection or misclassification in some cases. On one hand, we only collect the main RFI components with variance-ratio over 1 per cent, the classifier may omit some RFI. On the other hand, PCA may perform not so well while the data are seriously contaminated by complex and strong interference. We expect the classifier to work better when the data are more smooth and stable.

We will also test and apply our methods on the spectral-line data of FAST and on other telescopes.

## ACKNOWLEDGEMENTS

The authors thank the referee for careful review and suggestions, Axel Jessner and Michael Kramer for comments and suggestions to our scientific contents. This work was supported by the National SKA Program of China Number 2020SKA0120200, the CAS-MPG LEGACY project, the National Key R&D Program of China Numbers 2017YFA0402600 and 2019YFB1312704, and the National Natural Science Foundation of China Numbers 11873067, 12041303, and 12041301. This work was also supported by the Open Project Program of the Key Laboratory of FAST, NAOC, Chinese Academy of Sciences, and the CAS Key Laboratory of FAST, National Astronomical Observatories, Chinese Academy of Sciences, and the Cultivation Project for FAST Scientific Payoff and Research Achievement of CAMS-CAS.

## DATA AVAILABILITY

We share the available analysing codes and the data for verifying the methods on the website: http://paperdata.china-vo.org/maoyuan/RFI_Classification/data_RFI.zip.

The specific published data includes: (1) A data array set we used in the *One-day RFI-identification Experiment*; (2) A sample of the data set (presented in Fig. 8) produced by the RFI classifier; (3) The data we used to measure the classification precision and recall in Section 3.3.2. We shared on this China-VO website the programs for producing and analysing the simulated RFI. Because all the simulated RFI are artificially generated, we only publish the code instead of the generated data. We also make public our classifier code on http://https://github.com/Yuanao/RFI-Classification.

The whole raw observational data are difficult to share due to its size (up to several TB) and also protected under the FAST commission data policy and it may be released later through the official channels of the telescope.

## APPENDIX A: IMPROVEMENTS OF THE RFI CLASSIFICATION

In this appendix, we present three improvements of our methods to help reduce classification mixtures.

Our classier can extract and classify most RFI correctly, but we also find few exceptions. These exceptions could cause misclassifications. There are mainly three kinds of misclassifications, one is the mixture of components; one is the repetition of the same wideband pulse in multicomponents; and the last one is the misjudgment between periodic and noised-contaminated RFI that has no period. We will describe these misclassifications separately in the following parts, and introduce the measures we take to separate them as much as possible.

### A1 Improvements to the classification of narrowband RFI

In rare cases, some strong narrowband RFI (often present in individual channels) could not be cleanly decomposed and would contaminate multiple PCA components. We show an example of two of such channelized RFI in Fig. A1, they show up in two separate PCA components with the same temporal variability. However, these components are indeed dominated by two prominent channels: channel 1059 and channel 1224. Once we plot the time-series in each channel, we could see that these are two independent RFI, one (1059) is an impulsive RFI and the other (1024) is a recurrent one. Apparently, these two RFI are mixed in at least two different PCA components.

In such a situation, a basis's dominated frequency (or channel) is also a mixed channel set that contains different types of strong narrowband RFI. As to the reason of their occurrence in multicomponents, it is because regularized time-domain signal values alternate between positive and negative. Therefore, basis vectors with different positive and negative signs are needed to decompose these features. To prevent such strong narrowband RFI from contaminating multiple components, we sum the amplitude of all bases, and find any RFI channel that has a summed amplitude $>0.9$ dB (in relative intensity). Only those prominent and recurring RFI channels will have such a large summed amplitude. We then try to separate this strong RFI by keeping it only on one basis where its amplitude is at maximum. After removing this RFI frequency from all the other bases, we project the data on to the new bases.

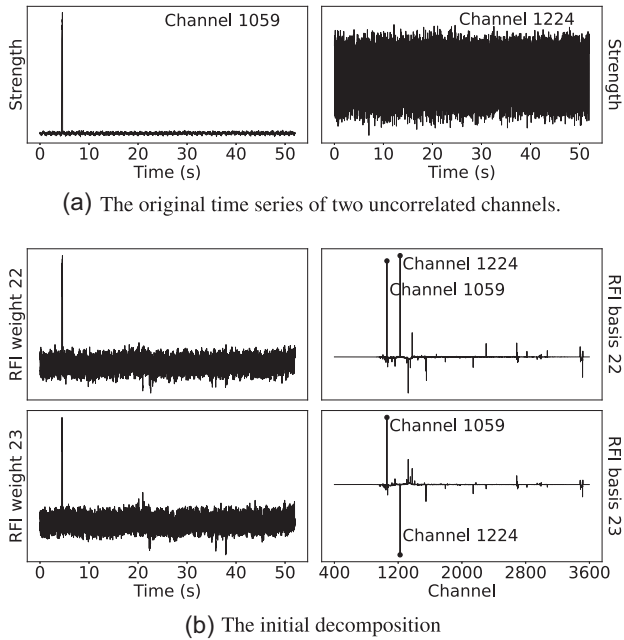

**Figure A1.** A decomposition result of a periodic and impulsive RFI from the original PCA operation. The top panel (a) shows the real temporal series of the two RFI on two unrelated channels. The bottom panel (b) is the 22nd and 23rd weight/basis derived by original PCA. In this case, channels 1059 and 1224 are mixed and simultaneously become dominant channels on both bases. Consequently, both the two weights show the impulsive and periodic feature which respectively belongs to channel 1059 and 1224.

Finally, we use the new bases to decompose the data and obtain the new weights. These strong channelized RFI are often well separated after these steps as shown in the case presented in Fig. A2. It shows that an improved PCA could separate the strong channelized RFI better than the generic ones, and thus help better categorize the RFI.

## A2 Improvements to the classification of wideband impulsive RFI

The aforementioned PCA process is effective at reducing the reappearance of narrowband RFI, but sometimes we also get broad-band impulsive RFI that cannot be confined to a single channel. In this case, we need to distinguish them from the above narrowband RFI. Based on real observations, we find the wideband impulses often manifest themselves in multiple components. Therefore, we classify any strong impulsive RFI as broad-band impulsive RFI if they repeat themselves in more than one components at different frequencies.

Fig. A3 shows such a wideband RFI example that appears at around 02:49:48. In this case, the wideband impulse repeats itself on several components different in frequencies (component 2 to 6). The re-identified wideband impulse is exhibited in the diagnostic plots, Fig. 8 (e.g. sub-figure b, as well as other narrowband pulses). In addition to this wideband interference, the components 17 and 18 of this figure also show another narrowband impulse. This pulse, appears at 02:49:25, is mixed with another interference. This is an example of mixed RFI components mentioned in the above Section A1. The improved decomposition result is shown in the sub-figure (b) of Fig. 8.

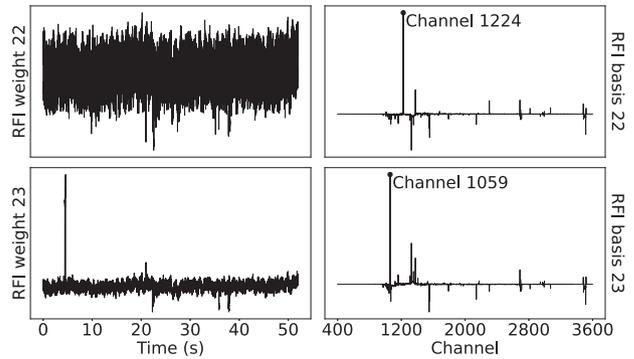

**Figure A2.** New RFI weights and bases after re-processing on PCA. As a comparison to the original PCA in Fig. A1, in this panel, channels 1059 and 1224 are disconnected on two components. As the projection, the weight 22 shows the periodic feature which belongs to channel 1224, and the weight 23 is impulsive RFI from channel 1059.

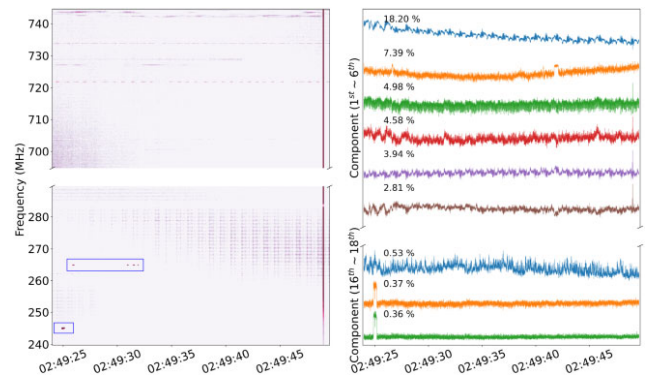

**Figure A3.** A presentation of the wideband impulsive RFI in raw data (left) and the decomposition result from original PCA (right). The left-hand panel shows the wideband impulse rises at 02:49:49. There are another two blue boxes in this panel, where indicate two narrowband impulsive events. One is a single-channel impulse at 02:49:25, 245.75 MHz, and another is a single-channel multi-impulses around 02:49:30, 265 MHz (i.e. the impulsive components in sub-figure b, Fig. 8). The right-hand panel is the decomposition of the original component analysis. It shows that the wideband impulse contaminated four main components (component 2 ∼ 6), and the narrowband impulse only appear on two minor components which is a misclassification case introduced above. The results in sub-figures (a) and (b) of Fig. 8 are the improved classification result of these cases.

## A3 Improvements to the classification of periodic RFI

Our PCA decomposition allows us to identify periodic signals in varying weights of the RFI components. We classify such signals as periodic RFI.

Generally, a periodic RFI could be easily identified through one of multiple harmonically related peaks in the Fourier transform of a time-series. However, FFT spectrum of some unstable temporal signals also have prominent low-frequency power that could be misinterpreted as peaks of periodic RFI. Therefore, a simple threshold method might be defective. We show two RFI with unstable temporal variation in Fig. A4 (the black lines on the left-hand panel). In most cases, such temporal variation is caused by the slow varying of signal strength during the observation time-scale of a data set.

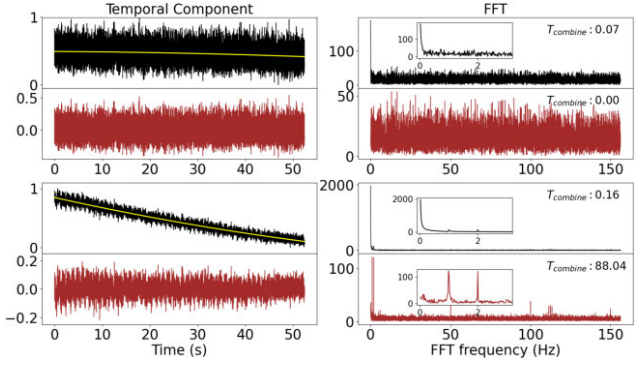

**Figure A4.** Two representative cases showing the process of removing the baseline of temporal varying signals. Left-hand panels are the weight time series of the extracted RFI components (black lines), the residual time-series (red lines) after quadratic baselines (yellow curves) are removed from the signal; Right-hand panel: the Fourier spectra of the time-series on the left. This figure shows how red-noise affects the low-frequency end in the Fourier spectra of two types of temporal RFI and how we could acquire clean spectra by first removing a quadratic low-frequency feature from the signals.

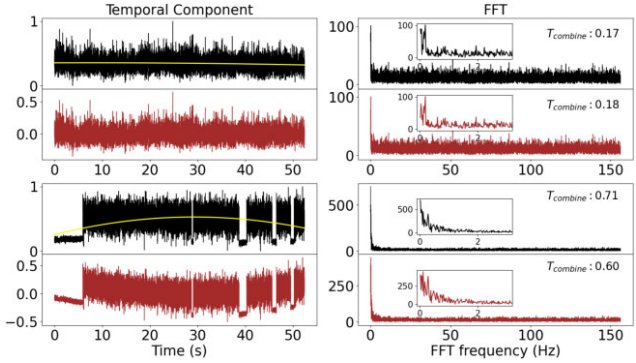

**Figure A5.** Two representative cases showing the process of how our classifier categorizes periodical and non-stationary RFI signals. Left-hand panels are the weight time-series and their baseline removing residues; Right-hand panel: the Fourier spectra of the series on the left. These two cases show that remarkable power in the Fourier power of non-stationary RFI still exits after baseline removing. We can correctly identify them through the combined threshold ($T_{\text{combine}} < 1$ Hz). A periodic RFI with $T_{\text{combine}} \sim 88$ Hz is shown in Fig. A4.

To prevent such temporal variation from contaminating real periodic signals, we remove quadratic polynomial fitted baselines from the temporal signals before we perform the FFT. We use a quadratic polynomial rather than a higher order polynomial because clipping the baseline fitted by a higher order polynomial may remove the authentic periodic fluctuations of the signals. We show two common cases of unstable time-series and their residuals after removing the fitted baseline in Fig. A4. The FFT spectra of both the raw signals and their cleaned residuals are also presented. In this figure, the black lines on the left side are the weights of PCA components and the red lines are the residuals after removing the fitted baseline (the yellow lines), which is overlapped on the weights. We plot four sub-panels for each case. 1, the top four sub-panels show the case that the remarkable power of an unstable signal's FFT spectrum is eliminated by removing the floating baseline. 2, the bottom four sub-panels show the case that the peaks of periodic RFI's FFT spectrum become clear after removing the remarkable low-frequency power caused by a slowly varying baseline.

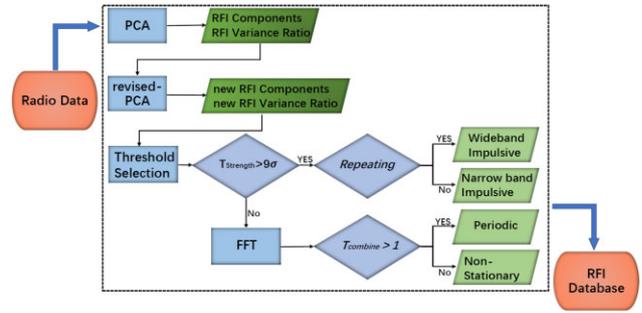

**Figure A6.** The data-processing procedures of RFI classifier. On the left is a raw radio data set as the input. The contents in the dashed frame are processes of how the classifier extracts and classifies RFI. The final output contains the RFI catalogue, RFI channel list, and necessary graphic illustrations.

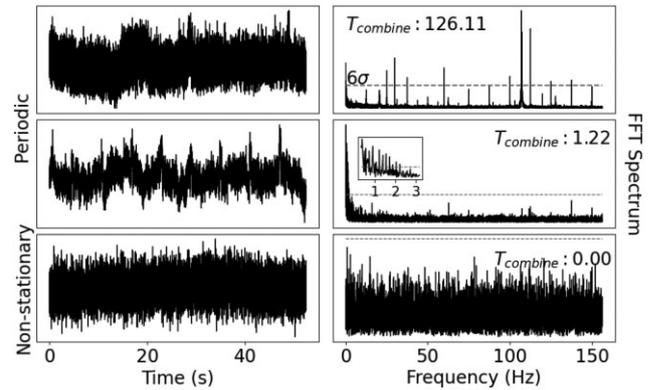

**Figure A7.** A presentation of periodic and non-stationary RFI components along with their FFT spectral. The RFI components here are from the periodic and non-stationary RFI components in Fig. 8.

A threshold selection can identify most artificial periodic RFI after an FFT performance to the baseline removed temporal components. However, baseline removing is not effective for all non-stationary RFI. Sometimes a temporal varying RFI also could be caused by random sudden fluctuations. In such cases, baseline removing will be defective. We show two such cases in Fig. A5. In this figure, both the two cases still have prominent power at the lower frequency end in the Fourier spectrum after removing the baseline.

We developed a combined threshold, $T_{\text{combine}}$, to distinguish the periodic RFI from non-stationary RFI better. Because peaks (or prominent power) on the Fourier spectrum of non-stationary RFI usually continuously distributes at the lower Fourier frequency band. While the spectrum peaks of artificial periodic RFI in pulsar data tend to locate at higher Fourier frequency and often discretely distribute on the spectrum due to the harmonics of period. On this consideration, we define the $T_{\text{combine}}$ (in Hz) as the sum of two specific values to measure the distribution of the remarkable power on the FFT spectrum. These two values are $\overline{f}_{\text{pulse}}$ (in Hz) and $\sigma(f_{\text{pulse}})$ (in Hz). The first criteria measure how much the peaks in Fourier spectrum locates at the high-frequency side. The second one is to measure whether the peaks are discretely distributed.

Specifically, $\overline{f}_{pulse}$ is the mean value of the recurrent frequencies $f_{pulse}$. $f_{pulse}$ here is the value of index frequencies (in Hz) with Fourier power $>6\sigma$ in the FFT spectrum. For common periodic RFI, $\overline{f}_{pulse}$ is remarkable through detecting the apparent pulses in the FFT spectrum.

$\sigma(f_{pulse})$, the variance of $f_{pulse}$, we define it to judge whether the remarkable frequency at the FFT spectrum is continuous or discrete. As the bottom right panel in Fig. A4 shown, the frequency at the spectrum with significant power is continuous for non-stationary series, but discrete for periodic RFI. Combining $\sigma(f_{pulse})$ with $\overline{f}_{pulse}$, we identify periodic RFI when $T_{combine} = \overline{f}_{pulse} + \sigma(f_{pulse}) > 1$ Hz. It should be noted that, we use the designed threshold on the Fourier spectrum of baseline removed series for the purpose of identifying periodic RFI more accurately, as the bottom panel in Fig. A4 shows, the $T_{combine}$ is valid after removing the decreasing baseline.

Here, we choose 1 Hz as the specific threshold is based on several factors: (1) the time-scale for an astronomical object that drifts across our beam is about 6–12 s; (2) the amplifiers in the telescope signal path vary in a time-scale of seconds that leads to red noise $<1$ Hz. Therefore, we chose 1 Hz to identify periodic RFI from other non-astronomical interference generating sources.

After some improvements to the RFI classifier, misclassification cases are reduced. The final improved data-processing procedure of the classifier is illustrated in Fig. A6.

To summarize, we designed an ensemble threshold $T_{combine}$ to better identify periodical RFI against red-noise, as shown in Fig. A7.

This paper has been typeset from a TEX/LATEX file prepared by the author.